\documentclass[aps,pra,amsfonts,amssymb,amsmath,twocolumn,showpacs]{revtex4}
\usepackage[dvips]{graphicx,epsfig}
\usepackage{ulem}
\def\ket#1{|#1\rangle}

\def\bracketi#1#2{\langle #1|#2 \rangle}

\def\ketbra#1#2{| #1 \rangle \langle #2 |}
\begin{document}
\title{Quantum non-demolition measurement of photon-arrival using an atom-cavity system}
\author{Kunihiro Kojima}
\email{kojima@qci.jst.go.jp}
\author{Akihisa Tomita}
\email{tomita@qci.jst.go.jp}
\affiliation{Quantum Computation and Information Project, ERATO-SORST, JST, Miyukigaoka 34 Ibaraki 305-8501, Japan}
\begin{abstract}
A simple and efficient quantum non-demolition measurement (QND) scheme is proposed in which the arrival of a signal photon is detected without affecting the qubit state. The proposed QND scheme functions even if the ancillary photon is replaced with weak light composed of vacuum and one-photon states. Although the detection scheme is designed for entanglement sharing applications, it is also suitable for general purification of a single photon state.
\end{abstract}
\maketitle
\section{Introduction}
Social needs for secure communications to prevent eavesdropping, impersonation, falsification, and denial of service have increased dramatically in recent years accompanying the expansion of the service industry on the internet.
Communication protocols based on quantum entanglement with non-local correlation have been the focus of intensive research as a possible means of providing secure communications \cite{ekert, gisin}.
Such protocols often require prior sharing of entangled photons \cite{kokone} or entangled atomic systems \cite{cabrillo, feng, chimczak, loock, childress} among more than two nodes. However, there is a practical difficulty in sharing entangled photons between distal nodes, since the photons are usually transmitted over lossy communication channels \cite{gisin, bras}. In such situations, photons entangled at the input of the channel can readily become a useless mixture of vacuum and photons at the output. To address this problem, it is necessary to purify the final mixture by removing the vacuum component \cite{bennett, briegel}. Quantum non-demolition (QND) measurements, by which the arrival of a signal photon is detected without affecting the qubit state (encoded into the polarization mode), may be very suitable for this purpose \cite{kokone, kok}. In this study, a new scheme for QND measurement that can be applied in cases involving the mixture of vacuum and photons is proposed.

\begin{figure}[htbp]
\begin{picture}(0,0)
\put(-120,-10){(a)}
\put(-120,-55){(b)}
\put(-120,-170){(c)}
\end{picture}
\begin{center} 
\includegraphics[width=5.0cm]{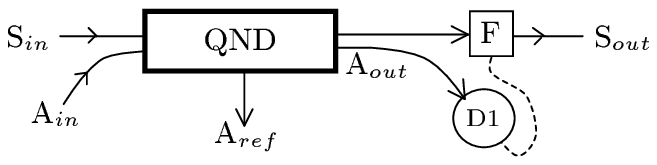}\\

\includegraphics[width=7.0cm]{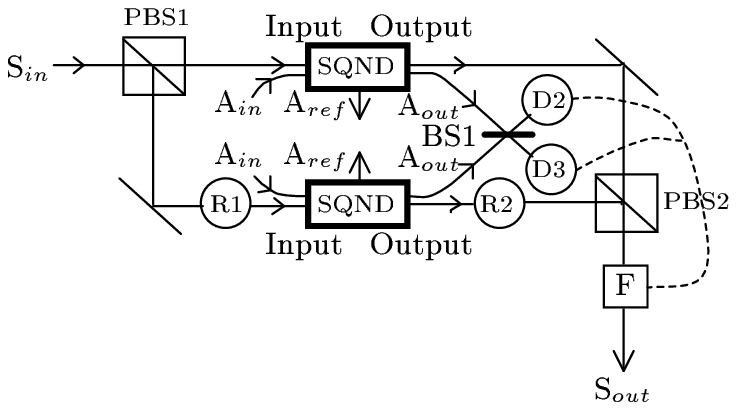}\\

\includegraphics[width=6.7cm]{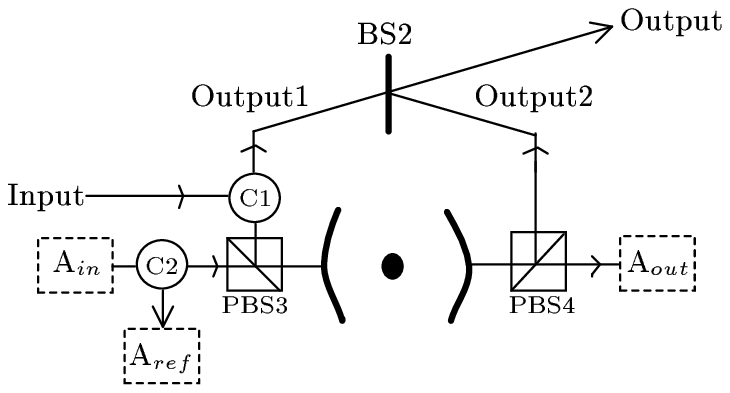}
\caption{\label{fig:qnd} \scriptsize (a) Schematic representation of QND measurement of a signal photon. (b) Proposed QND measurement scheme. (c) Proposed implementation of SQND for a single-mode polarized photon.\\ \({\rm S}_{in/out}\): Input/output channel of signal photon; \({\rm A}_{in/out/ref}\): Input/output/reflection channel of ancillary photon; {\rm F}: Filter; {\rm BS}: Half beam splitter; {\rm R}: Polarization rotator (changes the polarization of the input photon to the orthogonal polarization); {\rm C}: Circulator; {\rm D}: Photo-detector; {\rm PBS}: Polarization beam splitter}
\end{center} 
\end{figure}

Figure~\ref{fig:qnd}(a) shows the general process of QND detection. In this scheme, an ancillary photon prepared at the ancillary input \({\rm A}_{in}\) is transmitted and the photon is detected at the detector \({\rm D1}\) of the ancillary output \({\rm A}_{out}\) only when a signal photon appears at the signal input \({\rm S}_{in}\). The mixed state composed of vacuum and one-photon states at \({\rm S}_{in}\) is thus purified into a one-photon state at the output port \({\rm S}_{out}\) after filtering by filter {\rm F}, which allows the signal photon to pass through the filter only when the detector {\rm D1} detects the ancillary photon. When there is no photon at \({\rm S}_{in}\), ancillary photons appear at the reflection port \({\rm A}_{ref}\).

It is necessary to compose the corresponding realizations for the above scheme such that the qubit state of the signal photon is unchanged. Figure~\ref{fig:qnd}(b) shows the proposed QND scheme for the case that the qubit state is encoded into the polarization mode of the signal photon. When the signal photon appears at \({\rm S}_{in}\), the photon is transmitted or reflected at the polarization beam splitter {\rm PBS1} depending on the polarization mode. The polarization of the transmitted photon is orthogonal to that of the reflected photon. After passing through {\rm PBS1}, the wave packet of the signal photon is described by the superposition of the transmitted and reflected wave-packet components. Each component then enters into the QND with a single polarization mode ({\rm SQND}). The elements {\rm SQND} on each path are identical.

The process in {\rm SQND} should be the same as that for {\rm QND} in Fig.~\ref{fig:qnd}(a) except in two aspects. To maintain coherence between the transmitted and reflected components of the signal photon, the transmitted wave-packet components of the ancillary photons at each ancillary output \({\rm A}_{out}\) should be combined at the half beam splitter {\rm BS1} before detection by {\rm D2} and {\rm D3} (Fig.~\ref{fig:qnd}(b)) in order to erase information on the path taken by the signal photon. For the same purpose, the input state at each ancillary input port \({\rm A}_{in}\) in Fig.~\ref{fig:qnd}(b) must be a superposition of vacuum and one-photon states. In particular, when the input state at each \({\rm A}_{in}\) is a one-photon state, the wave-packet component at each ancillary reflection port \({\rm A}_{ref}\) must be detected by the same procedure as for the wave-packet component at each ancillary output port \({\rm A}_{out}\). The signal component at each {\rm output} shown in Fig.~\ref{fig:qnd}(b) is recombined by {\rm PBS2} after or before detection for ancillary photons. The signal photon thus passes through the filter {\rm F} only when an ancillary photon is detected at {\rm D2} or {\rm D3}.

The functionality of {\rm SQND} has been proposed and demonstrated based on a \(\chi^{(3)}\) material for modulating the phase of the ancillary photon only when a signal photon passes through the material \cite{chithreeqnd, grangier}. The phase change is then detected by a single-photon self-interference measurement after separating the ancillary photon from the signal photon, which are orthogonally polarized, by {\rm PBS}. However, phase modulation by a single photon is very small \cite{kok}, and the interference related to phase modulation will be suppressed due to entanglement between the signal photon and the ancillary photon, which changes the pulse shape of the ancillary photon. The self-interference effect is thus reduced, degrading the efficiency of {\rm SQND}. To avoid the use of weak \(\chi^{(3)}\) nonlinearities, a single-photon QND device composed of linear optics and projective measurements has been proposed. However, such a scheme requires strict mode-matching between the signal photon and the probe photon as one of a maximally entangled photon pair \cite{kok, jacob}.

The QND measurement scheme is implemented in the present study by a simple and efficient method involving a two-sided atom-cavity system consisting of two identical mirrors and an atom coupled with a single mode of a cavity. Figure~\ref{fig:qnd}(c) shows a schematic of the proposed implementation for {\rm SQND}. The polarizations of the signal and ancillary photons are orthogonal, and the two photons are combined by the polarization beam splitter {\rm PBS3} and directed to the two-sided atom-cavity system. The solid circle in Fig.~\ref{fig:qnd}(c) represent the intra-cavity atom, which interacts with photons via the cavity mirrors. The ancillary photons are considered to be totally reflected by the atom cavity when there is no signal photon at the input. This is realizable as suggested by the experiments of Turchette and coworkers \cite{turchetteone,turchettetwo}. When a signal photon arrives at the input, reflection of ancillary photons at the atom cavity is suppressed by saturation of the atomic transition due to the absorption of signal photons. Detection of the ancillary photon at {\rm D2} and {\rm D3} thus implies the arrival of the signal photon.

In the scheme shown in Fig.~\ref{fig:qnd}(c), the circulator {\rm C1} plays the role of separating the signal photon reflected at the atom cavity from the photon at the input port. Likewise, {\rm C2} separates the ancillary photon reflected at the atom cavity from the photon at the ancillary input port \({\rm A}_{in}\). {\rm PBS4} separates the signal photon transmitted at the atom cavity from the transmitted ancillary photon. The signal photon thus appears at output port 1 or 2, where the wave packet of the signal photon is described by the superposition of the wave-packet components of output ports 1 and 2. These components are then combined by the half beam splitter ({\rm BS2}) and the signal wave-packet appears at {\rm Output} when perfect mode-matching at {\rm BS2} is achieved. For the unachievable case, the wave-packet component of the signal photon leaks out on the opposite side of the {\rm Output} at {\rm BS2}. The leaked component will still be useful for QND as shown in Fig.~\ref{fig:qnd}(b) if the leaked components on each path are combined by {\rm PBS} and another signal output port is prepared.

The performance of the proposed QND scheme (Fig.~\ref{fig:qnd}(b)) is characterized in terms of efficiency and success probability, where the efficiency is defined as the probability that the signal photon appears at output 1 or 2 (Fig.~\ref{fig:qnd}(c)) when the ancillary photon is detected, while the success probability is defined as the probability that the ancillary photon is detected when the signal photon appears at output port 1 or 2. To estimate these quantities, the responses of the two-sided atom-cavity system for one- and two-photon input was analyzed considering a range of pulse durations for the input photons. Useful conditions for the QND are also examined. The pulse shape of the output signal photon after the time-resolved detection of the ancillary photon is analyzed qualitatively, since information on the pulse shape is important for processing the signal photons with other photons. It is found to be possible to increase the efficiency by up to 100 \% by increasing the pulse duration of the ancillary photon. However, the success probability is decreased simultaneously to 0 \% in a trade-off relationship. The efficiency is maintained even if the ancillary photon is replaced with weak light described by the superposition of vacuum and one-photon states, although the success probability is reduced in such a case. These results suggest that the mode-matching between the input light and the cavity mode is less critical in the proposed scheme \cite{hans} than in the QND proposals based on interferometry \cite{chithreeqnd, kok, jacob}.

The remainder of this paper is organized as follows. In Sec.~II, the Hamiltonian for the two-sided atom-cavity system is presented and the corresponding model is introduced. The output state for the atom-cavity system is then derived for one-photon pulse input (Sec.~III), and the output state for two-photon input is obtained using the results for one-photon input (Sec.~IV). In Sec.~V, the performance of the proposed QND is examined quantitatively and the coherence of the output signal photons is analyzed qualitatively for the case that the influence of the dephasing in the QND process is negligible. The implementation of the proposed QND is then discussed in Sec.~VI.

\section{\label{sec:model} Theoretical Model}
\begin{figure}[ht]
\begin{center} 
\includegraphics[width=6cm]{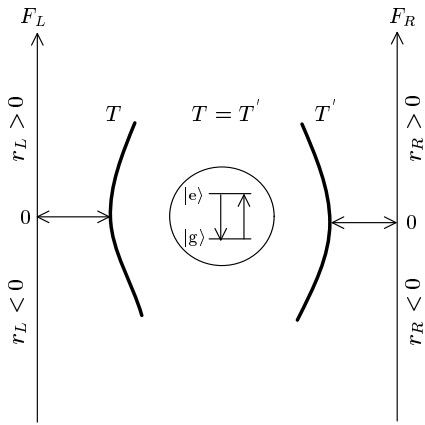}
\caption{\label{fig:model} \scriptsize Schematic representation of cavity geometry }
\end{center} 
\end{figure}

To analyze the responses of a two-sided cavity, in which a single two-level system couples with the single mode of the cavity, for one- and two-photon pulse input, a model of spatiotemporal propagation to and from the two-sided atom-cavity system is necessary. The proposed model is illustrated in Fig.~\ref{fig:model}. The cavity couples with the left-side field mode \(F_{L}\) and the right-side field mode \(F_{R}\) via the two mirrors, which have transmittance \(T\) and \(T^{`}\) (\(T = T^{'}\)). In the figure, {\rm g} and {\rm e} denote the ground and excited states of the single two-level atom. It is assumed that only one longitudinal and transversal mode is allowed in the cavity. The vertical arrow on the left side of the cavity (\(F_{L}\)) represents the radiative input (\(r_{L} < 0\)) and output (\(r_{L} > 0\)) fields at the cavity, where \(r_{L}\) corresponds to the spatial coordinate. The vertical arrow on the right side (\(F_{R}\)) similarly represents the input (\(r_{R} < 0\)) and output (\(r_{R} > 0\)) fields. The two-way arrows at the origin on the left and right arrows represent coupling of the atom-cavity system with the radiative fields \(F_{L}\) and \(F_{R}\).

The total Hamiltonian for this model is as follows.
\begin{eqnarray}
&& {} \hat{H} = \sum_{i=L,R} \left( \hat{H}_{F_{i}}+\hat{H}_{int F_{i}}\right)  + \hat{H}_{int ac} \label{eq:totalhamiltonian}
\end{eqnarray}
\begin{eqnarray}
\mbox{with} && {} \hat{H}_{F_{i}} = \int^{\infty}_{-\infty}dk\ \hbar c k_{i} \hat{b}_{F_{i}}^{\dagger}(k)\hat{b}_{F_{i}}(k) \nonumber \\
&& {} \hat{H}_{int F_{i}} = \int^{\infty}_{-\infty}dk\ i\hbar\sqrt{\frac{c \kappa}{\pi}} \left( \hat{b}_{F_{i}}^{\dagger}(k) \hat{a}-\hat{a}^{\dagger}\hat{b}_{F_{i}}(k) \right) \nonumber\\
&& {} \hat{H}_{int ac} = \hbar g \left( \hat{a}^{\dagger} \hat{\sigma}_{-}+\hat{\sigma}_{-}^{\dagger}\hat{a} \right) \nonumber 
\end{eqnarray}
where \(\hat{\sigma}_{-}=\ketbra{{\rm g}}{{\rm e}}\), and \(\hat{a}\) and \(\hat{b}_{F_{i}}(k)\) are the annihilation operators for the single mode of the cavity and the radiative field \(F_{i}\) (\(i=L,R\)), respectively. The single mode of the cavity is resonantly coupled with the atomic system and all the Hamiltonians have been formulated in a rotating frame defined by the transition frequency of the atomic system \(\omega_{0}\). The wave vector is likewise defined in the rotating frame, that is, \(k_{F_{i}}\) is defined relative to the resonant wave vector \(\omega_{0}/c\). The factor \(\sqrt{c\kappa/\pi}\) is the coupling constant between the single mode of the cavity and the radiative field, where \(\kappa\) is the cavity decay rate due solely to the coupling of the cavity mode with the radiative field \(F_{i}\). The factor \(g\) is the coupling constant between the cavity mode and the two-level system.

The response of the two-sided atom-cavity depends on the relative magnitude of the cavity decay rate \(\kappa\) with respect to the coupling constant \(g\). The bad-cavity regime characterized by \(\kappa \gg g\) \cite{kojima,badcavity} is assumed, as described below. 
\section{One-photon processes}
In the following calculations, it is assumed that the two-level system is in the ground state before the arrival of the input one-photon pulse from the left side of the cavity. The state of the field-atom-cavity system for one-photon processes can be expanded on the basis of the wavenumber eigenstates \(\ket{k_{L}}\) and \(\ket{k_{R}}\) of the radiative fields, the excited state of the two-level system \(\ket{ {\rm E}}\), and the cavity one-photon state \(\ket{{\rm C}}\). The state \(\ket{k_{L}}\) denotes a state with the atom in the ground state {\rm g}, the cavity mode \(a\) and all modes of the "R field" \(k_{R}\) in the vacuum state, and one mode of the "L field" \(k_{L}\) in the first excited state, with the remaining states being the vacuum state, i.e., \(\ket{k_{L}}=\ket{{\rm g},0_{a},1_{k_{L}},0_{k_{R}}}\). Likewise, \(\ket{k_{R}}=\ket{{\rm g},0_{a},0_{k_{L}},1_{k_{R}}}\), \(\ket{{\rm C}}=\ket{{\rm g},1_{a},0_{k_{L}},0_{k_{R}}}\), and \(\ket{{\rm E}}=\ket{{\rm e},0_{a},0_{k_{L}},0_{k_{R}}}\). The quantum state for the one-photon process can then be written as
\begin{eqnarray}
\ket{\Psi(t)} && {} = \Phi({\rm E};t)\ket{{\rm E}} + \Lambda({\rm C};t) \ket{{\rm C}} \nonumber \\
&& {} + \int dk_{L} \ \psi(k_{L};t)\ket{k_{L}} \nonumber \\
&& {} +\int dk_{R} \ \phi(k_{R};t)\ket{k_{R}} \label{eq:statebasis}
\end{eqnarray}
On these bases, the Hamiltonian given by eq.~(\ref{eq:totalhamiltonian}) can be expressed as
\begin{eqnarray}
&& {} \hat{H}_{1ph} =\hbar c\hat{k}_{L}+\hbar c\hat{k}_{R} \nonumber \\
&& {} + i\hbar\sqrt{\frac{c\kappa}{\pi}} \int^{\infty}_{-\infty} dk_{L} \left(\ketbra{k_{L}}{{\rm C}}-\ketbra{{\rm C}}{k_{L}}\right) \nonumber \\
&& {} + i\hbar\sqrt{\frac{c\kappa}{\pi}} \int^{\infty}_{-\infty} dk_{R} \left(\ketbra{k_{R}}{{\rm C}}-\ketbra{{\rm C}}{k_{R}}\right) \nonumber \\
&& {} + \hbar g \left(\ketbra{{\rm C}}{{\rm E}}+\ketbra{{\rm E}}{{\rm C}}\right) \nonumber \\
&& {} \mbox{where } \hat{k}_{j} = \int^{\infty}_{-\infty} dk_{j} \ \ketbra{k_{j}}{k_{j}}. \label{eq:onephotonhamiltonian}
\end{eqnarray}
The equations for the temporal evolution of the probability amplitudes \(\Phi({\rm E};t)\), \(\Lambda({\rm C};t)\), \(\psi(k_{L};t)\) and \(\phi(k_{R};t)\) can thus be obtained from the Schr\"odinger equation \(i \hbar d/dt \ket{\Psi(t)} = \hat{H} \ket{\Psi(t)}\) using eqs.~(\ref{eq:statebasis}) and (\ref{eq:onephotonhamiltonian}) as follows.
\begin{eqnarray}
&& \frac{d}{dt} \Phi({\rm E};t) = -ig \Lambda ({\rm C};t) \label{eq:excitedamp}\\
&& \frac{d}{dt} \Lambda({\rm C};t) = -ig \Phi ({\rm E};t) \nonumber \\
&& {} -\sqrt{\frac{c \kappa}{\pi}} \int dk_{L} \ \psi(k_{L};t) - \sqrt{\frac{c\kappa}{\pi}} \int dk_{R} \ \phi(k_{R};t) \nonumber \\
&& \label{eq:cavityamp}\\
&& \frac{d}{dt} \psi(k_{L};t) = -i k_{L}c \psi(k_{L};t) +\sqrt{\frac{c\kappa}{\pi}} \Lambda({\rm C};t) \label{eq:fieldone}\\
&& \frac{d}{dt} \phi(k_{R};t) = -i k_{R}c \phi(k_{R};t) + \sqrt{\frac{c\kappa}{\pi}} \Lambda({\rm C};t), \label{eq:fieldtwo}
\end{eqnarray}

The evolutions \(\psi(k_{L};t)\) and \(\phi(k_{R};t)\) can be obtained by integrating eqs.~(\ref{eq:fieldone}) and (\ref{eq:fieldtwo}):
\begin{eqnarray}
&& {} \psi(k_{L};t) = e^{-i k_{L}c \left(t-t_{i}\right)} \psi(k_{L};t_{i}) \nonumber \\
&& {} +\sqrt{\frac{c\kappa}{\pi}} \int^{t}_{t_{i}}dt^{'}\ e^{-ik_{L}c \left(t-t^{'}\right)} \Lambda({\rm C};t^{'}) \label{eq:fieldonesol}\\
&& {} \phi(k_{R};t) = e^{-i k_{R}c \left(t-t_{i}\right)} \phi(k_{R};t_{i}) \nonumber \\
&& {} +\sqrt{\frac{c\kappa}{\pi}} \int^{t}_{t_{i}}dt^{'}\ e^{-ik_{R}c \left(t-t^{'}\right)} \Lambda({\rm C};t^{'}), \label{eq:fieldtwosol}
\end{eqnarray}
where \(t_{i}\) is the initial time of the evolution. To describe the evolution in real space, the results of the integrations of eqs.~(\ref{eq:fieldonesol}) and (\ref{eq:fieldtwosol}) are subjected to Fourier transformation using
\begin{eqnarray}
&& {} \psi_{j}(r_{j};t) \equiv 
\begin{cases}
\frac{1}{\sqrt{2\pi}} \int^{\infty}_{-\infty}dk_{j} \ e^{ik_{j} \cdot r_{j}}\psi(k_{j};t) \mbox{ for } r_{j} < 0 \\
-\frac{1}{\sqrt{2\pi}} \int^{\infty}_{-\infty}dk_{j} \ e^{ik_{j} \cdot r_{j}}\psi(k_{j};t) \mbox{ for } r_{j} > 0,
\end{cases} \nonumber \\
&& \mbox{ where } j = L, R \label{eq:fouriertrans}
\end{eqnarray}
As the incoming field \(r_{j} < 0\) is discontinuously connected to the outgoing field \(r_{j} > 0\) via the mirror of the cavity, which changes the phase of the incoming field by \(\pi\), the phase of the incoming field amplitude is different from that of the outgoing amplitude by \(\pi\).

The real-space representation of the temporal evolution on the field \(F_{L}\) is then given by
\begin{eqnarray}
&& \psi_{L}(r_{L};t) \nonumber \\
&& = 
\begin{cases}
\psi_{L}(r_{L}-c(t-t_{i});t_{i}) \text{\ for $r_{L} < 0$} \\
-\psi_{L}(r_{L}-c(t-t_{i});t_{i}) \text{\ for $c(t-t_{i}) < r_{L}$} \\
-\psi_{L}(r_{L}-c(t-t_{i});t_{i}) - \sqrt{\frac{2\kappa}{c}} \Lambda({\rm C};t-\frac{r_{L}}{c}) \\ \text{\ \ \ \ \ \ \ \ \ \ \ \ \ \ \ \ \ \ \ \ \ \ \ \ \ \ \ \ \ \ \ \ \ for $0 < r_{L} < c(t-t_{i})$.}
\end{cases} \nonumber
\end{eqnarray}
\vspace{-0.5cm}
\begin{eqnarray}
&& \label{eq:realspaceampone}
\end{eqnarray}
The first case corresponds to the single-photon amplitude propagating on the incoming field \(r_{L} < 0\), the second case corresponds to reflection of the single-photon amplitude by the left mirror of the cavity and then propagation on the outgoing field \(r_{L} > 0\), and the third case consists of two parts; the component reflected by the left mirror, and the amplitude of a single photon re-emitted into the outgoing field \(r_{L} > 0\) after absorption by the cavity.

Likewise, the real-space representation of the temporal evolution on the field \(F_{R}\) reads as
\begin{eqnarray}
&& \phi_{R}(r_{R};t) \nonumber \\
&& =
\begin{cases}
\phi_{R}(r_{R}-c(t-t_{i});t_{i}) \text{\ \ for $r_{R} < 0$} \\
-\phi_{R}(r_{R}-c(t-t_{i});t_{i}) \text{\ \ for $c(t-t_{i}) < r_{R}$}\\
-\phi_{R}(r_{R}-c(t-t_{i});t_{i}) - \sqrt{\frac{2\kappa}{c}} \Lambda({\rm C};t-\frac{r_{R}}{c}) \\ \text{\ \ \ \ \ \ \ \ \ \ \ \ \ \ \ \ \ \ \ \ \ \ \ \ \ \ \ \ \ \ \ \ \ \ for $0 < r_{R} < c(t-t_{i})$.}
\end{cases} \nonumber
\end{eqnarray}
\vspace{-0.5cm}
\begin{eqnarray}
&& \label{eq:realspaceamptwo}
\end{eqnarray}
The temporal evolution of the cavity one-photon amplitude can be obtained by integrating eq.~(\ref{eq:cavityamp}) and using the Fourier transform (\ref{eq:fouriertrans}):
\begin{eqnarray}
\Lambda({\rm C};t) && {} = -ig \int^{t}_{t_{i}}dt^{'}\ e^{-2\kappa \left(t-t^{'} \right)} \Phi({\rm E};t^{'}) \nonumber \\
&& {} + e^{-2\kappa \left(t-t_{i} \right)} \Lambda({\rm C};t_{i}) \nonumber \\
&& {} -\sqrt{2 \kappa c} \int^{t}_{t_{i}}dt^{'}\ e^{-2\kappa \left(t-t^{'}\right)} \psi_{L}(-c(t^{'}-t_{i});t_{i}) \nonumber \\
&& {} -\sqrt{2 \kappa c} \int^{t}_{t_{i}}dt^{'}\ e^{-2\kappa \left(t-t^{'}\right)} \phi_{R}(-c(t^{'}-t_{i});t_{i}) \nonumber \\ 
&& {} \label{eq:cavitystateamp}
\end{eqnarray}
Since, in the present analysis, the atom-cavity system is in the ground state before the one-photon input pulse propagating on the field \(F_{L}\) arrives at the system, the cavity-state amplitude \(\Lambda({\rm C};t_{i})\) and the excited-state amplitude \(\Phi({\rm E};t_{i})\) at the initial time are zero, and the field amplitude \(\psi_{L}(r_{L};t_{i})\) is zero for the region \(r_{L}>0\). Moreover, it is assumed that the state of the field \(F_{R}\) is initially the vacuum state, that is, the field amplitude \(\phi_{R}(r_{R};t_{i})\) is zero. Under these assumptions, eqs.~(\ref{eq:realspaceamptwo}) and (\ref{eq:cavitystateamp}) can be reduced as follows.
\begin{eqnarray}
\Lambda({\rm C};t) && {} = -ig \int^{t}_{t_{i}}dt^{'}\ e^{-2\kappa \left(t-t^{'} \right)} \Phi({\rm E};t^{'}) \nonumber \\
&& {} -\sqrt{2 \kappa c} \int^{t}_{t_{i}}dt^{'}\ e^{-2\kappa \left(t-t^{'}\right)} \psi_{L}(-c(t^{'}-t_{i});t_{i}) \nonumber \\
&& {} \label{eq:simplecavityamp}
\end{eqnarray}
\begin{eqnarray}
&& \phi_{R}(r_{R};t) \nonumber \\
&& =
\begin{cases}
0 \text{\ \ \ \ \ \ \ \ \ \ \ \ \ \ \ \ for $r_{R} < 0$ or $c(t-t_{i}) < r_{R}$}\\
-\sqrt{\frac{2\kappa}{c}} \Lambda({\rm C};t-\frac{r_{F_{R}}}{c}) \text{\ \ \ \ for $0 < r_{R} < c(t-t_{i})$.}
\end{cases} \label{eq:simplefieldamp}
\end{eqnarray}

It is assumed above that the atom-cavity system is in the bad-cavity regime characterized by \(\kappa \gg g\). The cavity one-photon amplitude given by eq.~(\ref{eq:simplecavityamp}) can then be approximated as
\begin{eqnarray}
\Lambda({\rm C};t) && {} \simeq -i \frac{g}{2\kappa} \Phi({\rm E};t) \nonumber \\
&& {} -\sqrt{2 \kappa c} \int^{t}_{t_{i}}dt^{'}\ e^{-2\kappa \left(t-t^{'}\right)} \psi_{L}(-c(t^{'}-t_{i});t_{i}) \nonumber \\
&& {} \label{eq:appcavityamp}
\end{eqnarray} 
The excited-state amplitude \(\Phi({\rm E};t)\) can be obtained by integrating eq.~(\ref{eq:excitedamp}) and then substituting eq.~(\ref{eq:appcavityamp}), affording
\begin{eqnarray}
\Phi({\rm E};t) && {} \simeq ig \sqrt{2 \kappa c} \int^{t}_{t_{i}}dt^{''}\ e^{-\frac{\Gamma}{2}\left(t-t^{''}\right)} \nonumber \\
&& {} \times \int^{t^{''}}_{t_{i}}dt^{'} \ e^{-2\kappa \left(t^{''}-t^{'} \right)} \psi_{L}(-c(t^{'}-t_{i});t_{i}) \nonumber \\
\mbox{, where } && {} \Gamma = g^{2}/\kappa \label{eq:appexcitedamp}
\end{eqnarray}
The temporal evolution of the excited-state amplitude is explicitly dominated by the atomic dipole relaxation characterized by the rate \(\Gamma\) and is implicitly and effectively restricted by the cavity decay characterized by the rate \(\kappa\) and the input pulse duration. The reduction of that amplitude due to the cavity decay prevents efficient interaction between the input photon and the two-level system, which is the starting point of efficient nonlinear two-photon interaction. It is therefore assumed that the pulse duration of the input one-photon is much larger than the cavity decay time \(1/\kappa\). The excited-state amplitude (\ref{eq:appexcitedamp}) can then be approximated as
\begin{eqnarray}
&& {} \Phi({\rm E};t) \simeq i \sqrt{\frac{c \Gamma}{2}} \int^{t}_{t_{i}}dt^{''}\ e^{-\frac{\Gamma}{2}\left(t-t^{''}\right)} \psi_{L}(-c(t^{''}-t_{i});t_{i}) \nonumber \\
&& {} \ \ \ \ \ \ \ \ \ \ \ \ \ \ \ \ \ \ \ \ \ \ \ \ \ \ \ \ \ \ \ \ \ \ \ \ \ \ \ \ \ \ \mbox{ for } t-t_{i} \gg 1/\kappa \label{eq:effectiveexcitedamp}
\end{eqnarray}
Likewise, the cavity one-photon amplitude given by eq.~(\ref{eq:appcavityamp}) can be approximated with eq.~(\ref{eq:effectiveexcitedamp}) as
\begin{eqnarray}
\Lambda({\rm C};t) && {} \simeq \frac{\Gamma}{2g} \sqrt{\frac{c \Gamma}{2}} \int^{t}_{t_{i}}dt^{''}\ e^{-\frac{\Gamma}{2}\left(t-t^{''}\right)} \psi_{L}(-c(t^{''}-t_{i});t_{i}) \nonumber \\
&& {} -\sqrt{\frac{c}{2 \kappa}} \psi_{L}(-c(t-t_{i});t_{i}) \mbox{ for } t-t_{i} \gg 1/\kappa \label{eq:effectivecavityamp}
\end{eqnarray}

The effective field amplitude for \(\psi_{L}(r_{L};t)\) can be obtained by substituting eq.~(\ref{eq:effectivecavityamp}) into eq.~(\ref{eq:realspaceampone}), giving
\begin{eqnarray}
&& \psi_{L}(r_{L};t) \nonumber \\
&& \simeq
\begin{cases}
\psi_{L}(r_{L}-c(t-t_{i});t_{i}) = 0 \text{\ \ \ \ \ for $c(t-t_{i}) < r_{L}$}\\
\psi_{L}(r_{L}-c(t-t_{i});t_{i}) \text{\ \ \ \ \ for $r_{L} < 0$} \\
-\frac{\Gamma}{2}\int^{t}_{t_{i}}dt^{''}\ e^{-\frac{\Gamma}{2}\left(t-r_{L}/c-t^{''}\right)} \psi_{L}(-c(t^{''}-t_{i});t_{i}) \\ \text{\ \ \ \ \ \ \ \ \ \ \ \ \ \ \ \ \ \ \ \ \ \ \ \ \ \ \ \ \ \ \ \ \ for $0 < r_{L} < c(t-t_{i})$.}
\end{cases} \nonumber
\end{eqnarray}
\begin{eqnarray}
&& \label{eq:effectiverealspaceampone}
\end{eqnarray}
Here, the first case is equal to zero, since the field amplitude \(\psi_{L}(r_{L};t_{i})\) is assumed to be initially equal to zero for \(r_{L} > 0\). The second case corresponds to the incoming field amplitude at the atom-cavity system, and the third case corresponds to the field amplitude of the photon re-emitted by the intra-cavity atomic system.

The effective field amplitude for \(\phi_{R}(r_{R};t)\) can similarly be obtained as
\begin{eqnarray}
&& \phi_{R}(r_{R};t) \nonumber \\
&& \simeq
\begin{cases}
\phi_{R}(r_{R}-c(t-t_{i});t_{i}) = 0 \\ \text{\ \ \ \ \ \ \ \ \ \ \ \ \ \ \ \ \ \ \ \ \ \ \ \ \ \ for $r_{R} < 0$ or $c(t-t_{i}) < r_{R}$}\\
-\frac{\Gamma}{2} \int^{t}_{t_{i}}dt^{''}\ e^{-\frac{\Gamma}{2}\left(t-r_{R}/c-t^{''}\right)} \psi_{L}(-c(t^{''}-t_{i});t_{i}) \nonumber \\
+\psi_{L}(r_{R}-c(t-t_{i});t_{i}) \text{\ \ \ \ for $0 < r_{R} < c(t-t_{i})$.}
\end{cases}
\end{eqnarray}
\begin{eqnarray}
&& \label{eq:effectivesimplefieldamp}
\end{eqnarray}
Here, the first case is equal to zero according to the initial condition, and the second case represents the interference between the field amplitude of the transmitted photon without absorption by the atomic system and the field amplitude of the photon re-emitted by the atomic system.

To investigate the outgoing amplitudes \(\psi_{L}(r_{L}>0;t)\); \(\phi_{R}(r_{R}>0;t)\) for an arbitrary incoming amplitude under the above-mentioned initial conditions, it is convenient to represent the outgoing amplitudes as a matrix element of the evolution operator, as follows.
\begin{eqnarray}
\psi_{L}(r_{L};t) && {} = \bracketi{r_{L}}{\Psi(t)} \nonumber \\
                      && {} = \int^{\infty}_{-\infty}dr^{'}_{L} \  {\rm u}_{1ph}^{(L)}(r_{L},r^{'}_{L};t-t_{i}) \nonumber \\
&& {} \times \psi_{L}(r^{'}_{L};t_{i}) \label{eq:matrixrepresentl}
\end{eqnarray}
\begin{eqnarray}
\phi_{R}(r_{R};t) && {} = \bracketi{r_{R}}{\Psi(t)} \nonumber \\
                      && {} = \int^{\infty}_{-\infty}dr^{'}_{L} \  {\rm u}_{1ph}^{(R)}(r_{R},r^{'}_{L};t-t_{i}) \nonumber \\
&& {} \times \psi_{L}(r^{'}_{L};t_{i}) \label{eq:matrixrepresentr}
\end{eqnarray}
Here, \({\rm u}_{1ph}^{(L)}(r_{L},r^{'}_{L};t-t_{i})\) and \({\rm u}_{1ph}^{(R)}(r_{R},r^{'}_{L};t-t_{i})\) are the matrix elements of the evolution operator \(e^{-\frac{i}{\hbar}\hat{H}_{\rm 1 ph}}\), representing the transition probability amplitude from the state \(\ket{r^{'}_{L}}\) at time \(t_{i}\) to the state \(\ket{r_{LorR}}\) at time \(t\), where \(\ket{r_{j}} \equiv \frac{1}{\sqrt{2\pi}}\int^{\infty}_{-\infty}dk_{j} \ e^{-ik_{j}r_{j}} \ket{k_{j}}\) for \(j = L, R\). These matrix elements can be obtained approximately by comparing the results of eqs.~(\ref{eq:effectiverealspaceampone}), (\ref{eq:effectivesimplefieldamp}),  (\ref{eq:matrixrepresentl}), and (\ref{eq:matrixrepresentr}), as follows.
\begin{eqnarray}
{\rm u}_{1ph}^{(L)}(r_{L},r^{'}_{L};t-t_{i}) && \simeq {} {\rm u}_{\rm abs}^{(L)}(r_{L},r^{'}_{L};t-t_{i}) \label{eq:lphoton} \\
{\rm u}_{1ph}^{(R)}(r_{R},r^{'}_{L};t-t_{i}) && \simeq {} {\rm u}_{\rm trans}^{(R)}(r_{R},r^{'}_{L};t-t_{i}) \nonumber \\
&& {} + {\rm u}_{\rm abs}^{(R)}(r_{R},r^{'}_{L};t-t_{i}) \label{eq:onephotonprocess}
\end{eqnarray}
\begin{eqnarray}
\mbox{ with } && {} {\rm u}_{\rm trans}^{(R)}(r_{R},r^{'}_{L};t-t_{i}) = \delta \left(r_{R}-c(t-t_{i})-r^{'}_{L}\right) \nonumber
\end{eqnarray}
\begin{eqnarray}
\mbox{ and } && {} {\rm u}_{\rm abs}^{(j)}(r_{j},r^{'}_{L};t-t_{i}) = \nonumber \\
&& {}
\begin{cases}
-\frac{\Gamma}{2c} e^{-\frac{\Gamma}{2c} \left(c(t-t_{i}) + r^{'}_{L}-r_{j}\right)} \\ \mbox{\ \ \ \ for } 0 < r_{j} < c\left(t-t_{i}\right) + r^{'}_{L} \mbox{ and } r^{'}_{L} < 0\\
0 \mbox{\ \ for } r_{j} > c\left(t-t_{i}\right) + r^{'}_{L} \mbox{ or } r^{'}_{L} > 0.
\end{cases} \nonumber \\
&& {} \mbox{\ \ \ \ \ \ \ \ \ \ \ \ \ \ \ \ \ \ \ \ \ \ \ \ \ \ \ \ \ \ \ \ \ \ \ \ \ \ \ \ \ \ \ \ \ \ \ for } j=L, R. \nonumber \\
&& \label{eq:onephabs}
\end{eqnarray}
The component \({\rm u}_{\rm trans}\) is the transition component for the transmitted photon without absorption by the atomic system, while \({\rm u}_{\rm abs}\) is the transition component for the photon re-emitted by the atomic system.

\section{Two-photon processes}
\begin{figure}[ht]
\begin{center} 
\includegraphics[width=6cm]{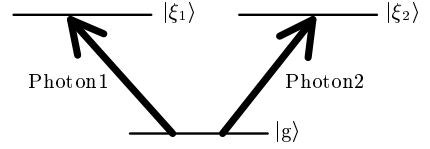}
\caption{\label{fig:vtype} \scriptsize V-type three-level system}
\end{center} 
\end{figure}

The interaction between two photons in the atom-cavity system is treated as follows. It is assumed that the two photons {\rm Photon1} and {\rm Photon2} are distinguishable, by the polarization mode in this case. The atomic system described by the theoretical model can thus be implemented as a V-type three-level system. In the V-type system, there are two excited states, \(\ket{\xi_{1}}\) and \(\ket{\xi_{2}}\), with orthogonal polarizations but sharing the same ground state \(\ket{{\rm g}}\) (see Fig.~\ref{fig:vtype}). The transition to the excited state \(\ket{\xi_{1 (2)}}\) is caused by {\rm Photon1 (2)}. According to this representation, the total Hamiltonian is given by
\begin{eqnarray}
&& {} \hat{H} = \sum_{i=L,R;j=1,2} \left( \hat{H}^{(j)}_{F_{i}}+\hat{H}^{(j)}_{int F_{i}}\right)  + \hat{H}^{(j)}_{int ac} \label{eq:exthamiltonian}\\
\mbox{with} && {} \hat{H}^{(j)}_{F_{i}} = \int^{\infty}_{-\infty}dk\ \hbar c k_{i} \hat{b}_{F_{ij}}^{\dagger}(k)\hat{b}_{F_{ij}}(k) \nonumber \\
&& {} \hat{H}^{(j)}_{int F_{i}} = \int^{\infty}_{-\infty}dk\ i\hbar\sqrt{\frac{c \kappa}{\pi}} \left( \hat{b}_{F_{ij}}^{\dagger}(k) \hat{a}_{j}-\hat{a}_{j}^{\dagger}\hat{b}_{F_{ij}}(k) \right) \nonumber\\
&& {} \hat{H}^{(j)}_{int ac} = \hbar g \left( \hat{a}_{j}^{\dagger} \hat{\sigma}^{(j)}_{-}+\hat{\sigma}_{-}^{\dagger (j)}\hat{a}_{j} \right) \nonumber 
\end{eqnarray}
where \(\hat{\sigma}^{(j)}_{-} = \ketbra{{\rm g}}{\xi_{j}}\), and \(\hat{a}_{j}\) and \(\hat{b}_{F_{ij}}(k)\) are the annihilation operators for the $j$th mode of the cavity and the radiative field \(F_{ij}\) (\(i=L,R\) and \(j=1,2\)), respectively. The other conditions are the same as in the theoretical model.

To extend the response function for the one-photon input given by eqs.~(\ref{eq:lphoton}) and (\ref{eq:onephotonprocess}) to the two-photon case, the state description for one-photon processes presented above must be extended to two-photon processes to afford the total Hamiltonian given by eq.~(\ref{eq:exthamiltonian}) for the distinguishable two-photon input involving {\rm Photon1} and {\rm Photon2}. The states for two-photon processes obtained by extending the state description for one-photon processes are as follows.
\begin{eqnarray}
&& \ket{{\rm C}_{1}} \otimes \ket{{\rm C}_{2}}, \ket{{\rm E}_{1}} \otimes \ket{{\rm C}_{2}} \nonumber \\
&& \ket{{\rm C}_{1}} \otimes \ket{{\rm E}_{2}}, \ket{k_{i1}} \otimes \ket{k_{i^{'}2}} \nonumber \\
&& \ket{{\rm C}_{1}} \otimes \ket{k_{i2}}, \ket{k_{i1}} \otimes \ket{{\rm C}_{2}} \nonumber \\
&& \ket{{\rm E}_{1}} \otimes \ket{k_{i2}}, \ket{k_{i1}} \otimes \ket{{\rm E}_{2}}, \ket{{\rm E}_{1}} \otimes \ket{{\rm E}_{2}} \label{eq:twophotondescription}
\end{eqnarray}
Here, \(i,i^{'}=L,R\). The state \(\ket{k_{Lj}}\) denotes a state with the atom in the ground state \({\rm g}_{j}\), the cavity mode \(a_{j}\) and all modes of the "Rj field" \(k_{Rj}\) in the vacuum state, and one mode of the "Lj field" \(k_{Lj}\) in the first excited state, with the remaining states being the vacuum state, i.e., \(\ket{k_{Lj}}=\ket{{\rm g}_{j},0_{a_{j}},1_{k_{Lj}},0_{k_{Rj}}}\). Likewise, \(\ket{k_{Rj}}=\ket{{\rm g}_{j},0_{a_{j}},0_{k_{Lj}},1_{k_{Rj}}}\), \(\ket{{\rm C}_{j}}=\ket{{\rm g}_{j},1_{a_{j}},0_{k_{Lj}},0_{k_{Rj}}}\), and \(\ket{{\rm E}_{j}}=\ket{{\rm e}_{j},0_{a_{j}},0_{k_{Lj}},0_{k_{Rj}}}\) for \(j=1,2\).

Denoting the states \(\ket{{\rm g}}\), \(\ket{\xi_{1}}\), and \(\ket{\xi_{2}}\) of the V-type system by \(\ket{{\rm g}_{1},{\rm g}_{2}}\), \(\ket{{\rm e}_{1},{\rm g}_{2}}\), and \(\ket{{\rm g}_{1},{\rm e}_{2}}\), the interaction Hamiltonian \(\hat{H}^{(j)}_{intac}\) given in eq.~(\ref{eq:exthamiltonian}) can be rewritten for the interaction with two distinguishable photons ({\rm Photon1} and {\rm Photon2}) as
\begin{eqnarray}
&&  \sum_{j=1,2} \ \hat{H}^{(j)}_{intac} \nonumber \\
&& {} = \hbar g \sum_{i=L,R}\ \left(\ketbra{{\rm C}_{1}}{{\rm E}_{1}}+\ketbra{{\rm E}_{1}}{{\rm C}_{1}}\right) \nonumber \\
&& \ \ \ \ \ \ \ \ \ \ \ \ \ \otimes \left(\ketbra{{\rm C}_{2}}{{\rm C}_{2}} + \int^{\infty}_{-\infty} dk_{i2} \ \ketbra{k_{i2}}{k_{i2}}\right) \nonumber \\
&& {} \ \ \ \ \ \ \ \ \ \ \ \ + \left(\ketbra{{\rm C}_{1}}{{\rm C}_{1}} + \int^{\infty}_{-\infty} dk_{i1} \ \ketbra{k_{i1}}{k_{i1}}\right) \nonumber \\
&& \ \ \ \ \ \ \ \ \ \ \ \ \ \otimes \left(\ketbra{{\rm C}_{2}}{{\rm E}_{2}}+\ketbra{{\rm E}_{2}}{{\rm C}_{2}}\right). \label{eq:intac}
\end{eqnarray}
See appendix A for the derivation of eq.~(\ref{eq:intac})). To facilitate formulation of the matrix element of temporal evolution for two-photon processes, the Hamiltonian given by eq.~(\ref{eq:intac}) is further divided into a linear term and a nonlinear term as follows.
\begin{eqnarray}
&& \hat{H}_{intac} = \hat{H}^{{\rm lin}}_{intac} + \hat{H}^{{\rm Nonlin}}_{intac}\nonumber \\,
&& \mbox{where } \hat{H}^{{\rm lin}}_{intac} = \hbar g \left( \left(\ketbra{{\rm C}_{1}}{{\rm E}_{1}}+\ketbra{{\rm E}_{1}}{{\rm C}_{1}}\right) \otimes \hat{I}^{(2)}_{1ph} \right. \nonumber \\
&& {} \ \ \ \ \ \ \ \ \ \ \ \ \ \ \ \ \ \ \ \ \left. + \hat{I}^{(1)}_{1ph} \otimes \left(\ketbra{{\rm C}_{2}}{{\rm E}_{2}}+\ketbra{{\rm E}_{2}}{{\rm C}_{2}}\right) \right) \nonumber \\
&& \label{eq:intaclin} \\
&& {} \hat{H}^{{\rm Nonlin}}_{intac} = - \hbar g \left( \left(\ketbra{{\rm C}_{1}}{{\rm E}_{1}}+\ketbra{{\rm E}_{1}}{{\rm C}_{1}}\right) \otimes \ketbra{{\rm E}_{2}}{{\rm E}_{2}} \right. \nonumber \\
&& {} \ \ \ \ \ \ \ \ \ \ \ \ \ \left. + \ketbra{{\rm E}_{1}}{{\rm E}_{1}} \otimes \left(\ketbra{{\rm C}_{2}}{{\rm E}_{2}}+\ketbra{{\rm E}_{2}}{{\rm C}_{2}}\right) \right) \nonumber \\
&& \label{eq:intacnonlin} \\
&& {} \mbox{with } \hat{I}^{(j)}_{1ph} = \sum_{i=L, R} ( \int^{\infty}_{-\infty} dk_{ij} \ \ketbra{k_{ij}}{k_{ij}} + \ketbra{{\rm C}_{j}}{{\rm C}_{j}} \nonumber \\
&& {} \ \ \ \ \ \ \ \ \ \ \ \ \ + \ketbra{{\rm E}_{j}}{{\rm E}_{j}} ) \nonumber
\end{eqnarray}
The linear term given by eq.~(\ref{eq:intaclin}) describes the dynamics of the two photons {\rm Photon1} and {\rm Photon2}, which are absorbed and emitted independently. The linear Hamiltonian includes transitions to the state \(\ket{{\rm E_{1}},{\rm E_{2}}}\), where both photons are absorbed by the atomic system. This transition is impossible in a two-level system (a V-type three-level system is considered here). The nonlinear term given by eq.~(\ref{eq:intacnonlin}) suppresses transitions to the state \(\ket{{\rm E_{1}},{\rm E_{2}}}\).

The total Hamiltonian given by eq.~(\ref{eq:exthamiltonian}) for two-photon processes is thus given by 
\begin{eqnarray}
\hat{H}_{2ph} &=& {} \hat{H}^{\rm lin} + \hat{H}^{\rm Nonlin}.
\end{eqnarray}

\begin{eqnarray}
&& \hat{H}^{\rm lin} = {} \hat{H}^{(1)}_{1ph} \otimes \hat{I}^{(2)}_{1ph} + \hat{I}^{(1)}_{1ph} \otimes \hat{H}^{(2)}_{1ph}, \label{eq:linearhamiltonian} \\
&& \hat{H}^{\rm Nonlin} = {} - \left(\hat{H}^{(1)}_{intac} \otimes \ketbra{{\rm E_{2}}}{{\rm E_{2}}} + \ketbra{{\rm E_{1}}}{{\rm E_{1}}} \otimes \hat{H}^{(2)}_{intac} \right) \nonumber \\
&& {} \label{eq:nonlinearhamiltonian}
\end{eqnarray}
\begin{eqnarray}
&& \mbox{ where } \hat{H}^{(j)}_{1ph} = \hat{H}^{(j)}_{intac} + \sum_{i=L, R} \hbar c \hat{k}^{(ij)} + \hat{H}^{(ij)}_{intfc} \nonumber \\
&& {} \ \ \ \ \ \ \ \ \ \ \ \ \ + \ketbra{{\rm E}_{j}}{{\rm E}_{j}} ) \nonumber \\
&& \mbox{ with } \hat{k}^{(ij)} = \int^{\infty}_{-\infty} dk_{ij} \ k_{ij} \ketbra{k_{ij}}{k_{ij}} \nonumber
\end{eqnarray}
\begin{eqnarray}
&& {} \mbox{ } \hat{H}^{(ij)}_{intfc} = i\hbar\sqrt{\frac{c\kappa}{\pi}} \int^{\infty}_{-\infty} dk_{ij} \left(\ketbra{k_{ij}}{{\rm C}_{j}}-\ketbra{{\rm C}_{j}}{k_{ij}}\right), \nonumber \\
&& {} \mbox{ and } \hat{H}^{(j)}_{intac} = \hbar g \left(\ketbra{{\rm C}_{j}}{{\rm E}_{j}}+\ketbra{{\rm E}_{j}}{{\rm C}_{j}}\right) \nonumber
\end{eqnarray}
where the indices \(i = L, R\) distinguish the left-side field of the two-sided cavity from the right-side field, and the indices \(j = 1, 2\) distinguish the two photons and the two excited states.

As the temporal evolution described by \(\hat{H}^{\rm lin}\) is composed of the evolution of a single photon, the corresponding matrix element of the temporal-evolution operator can be expressed as the product of the individual single-photon matrix elements given by eqs.~(\ref{eq:lphoton}) and (\ref{eq:onephotonprocess}), i.e., 
\begin{eqnarray}
&& {} {\rm u}^{{\rm lin} (jk)}_{2ph}(r_{j1},r_{k2};r^{'}_{L1},r^{'}_{L2};t-t_{i}) \nonumber \\
&& {} \ \ \ = {\rm u}_{1ph}^{(j)}(r_{j1};r^{'}_{L1};t-t_{i}) \cdot {\rm u}_{1ph}^{(k)}(r_{k2};r^{'}_{L2};t-t_{i}) \nonumber \\
&& {} \\
&& {} \mbox{ \ \ \ \ \ \ \ \ \ \ \ \ \ \ \ \ \ \ \ \ \ \ \ \ \ \ \ \ \ \ \ \ \ \ \ \ \ \ \ \ \ \ for } j, k = L, R. \nonumber
\end{eqnarray}
The components of reemission from the state \(\ket{{\rm E_{1}},{\rm E_{2}}}\) can be effectively described by
\begin{eqnarray}
&& {} {\rm u}_{\rm abs}^{(j)}(r_{j1},r^{'}_{L1};t-t_{i}) \cdot {\rm u}_{\rm abs}^{(k)}(r_{k2},r^{'}_{L2};t-t_{i}) \label{eq:nonlinterm} \\
&& {} \ \ \ \ \mbox{ for } 0 < r_{j1}, r_{k2} < c\left(t-t_{i}\right) + \mbox{{\rm Min}} \left[r^{'}_{L1}, r^{'}_{L2}\right] \nonumber
\end{eqnarray}
using eq.~(\ref{eq:onephabs}). These reemission components refer to the process in which the two photons are absorbed and reemitted simultaneously by an atom. However, the nonlinear term (\ref{eq:nonlinearhamiltonian}) eliminates these components \cite{kojima}. The total matrix element for two-photon processes can thus be described by
\begin{eqnarray}
&& {} {\rm u}^{(jk)}_{2ph}(r_{j1},r_{k2};r^{'}_{L1},r^{'}_{L2};t-t_{i}) \nonumber \\
&& {} \ \ \ = {\rm u}^{{\rm lin} (jk)}_{2ph}(r_{j1},r_{k2};r^{'}_{L1},r^{'}_{L2};t-t_{i}) \nonumber \\
&& {} \ \ \ + {\rm u}^{{\rm Nonlin} (jk)}_{2ph}(r_{j1},r_{k2};r^{'}_{L1},r^{'}_{L2};t-t_{i}), \label{eq:twophotonmatrixelement} \\
&& {} \mbox{ where } {\rm u}^{{\rm Nonlin} (jk)}_{2ph}(r_{j1},r_{k2};r^{'}_{L1},r^{'}_{L2};t-t_{i}) \simeq - \left( \mbox{ eq.~(\ref{eq:nonlinterm})} \right). \nonumber
\end{eqnarray}
The output wave-function on the left-side and right-side output fields can then be expressed as
\begin{eqnarray}
&& \Psi_{jk}(r_{j1},r_{k2};t) \nonumber \\
&& {} = \int^{\infty}_{-\infty} dr^{'}_{L1}dr^{'}_{L2} \ {\rm u}^{(jk)}_{2ph}(r_{j1},r_{k2};r^{'}_{L1},r^{'}_{L2};t-t_{i}) \nonumber \\
&& {} \ \ \ \ \times \Psi_{LL}(r^{'}_{L1},r^{'}_{L2};t_{i}) \label{eq:twophotonoutput} \\
&& {} \mbox{\ \ \ \ \ \ \ \ \ \ \ \ \ \ \ \ \ \ \ \ \ \ \ \ \ \ \ \ \ \ \ \ \ \ \ \ \ \ \ \ \ \ \ \ for } j,k = L, R. \nonumber
\end{eqnarray}
The output wave-function describes the far-field state of the photons after interaction with the atom-cavity system. In general, a two-photon wave-function propagating in free space is given by \(\Psi_{jk}(r_{j1},r_{k2};t) = \Psi_{jk}(r_{j1}-ct,r_{k2}-ct)\). The results of eq. (\ref{eq:twophotonmatrixelement}) and (\ref{eq:twophotonoutput}) can therefore be simplified by transformation to a moving coordinate system, i.e.,
\begin{eqnarray}
&& {} r_{j1}-ct = x_{j1} \nonumber \\
&& {} r_{j2}-ct = x_{j2} \nonumber \\
&& {} r^{'}_{L1}-ct_{i} = x^{'}_{L1}\nonumber \\
&& {} r^{'}_{L2}-ct_{i} = x^{'}_{L2} \nonumber \\
&& {} \mbox{\ \ \ \ \ \ \ \ \ \ \ \ \ \ \ \ \ \ \ \ \ \ \ \ \ \ \ \ \ \ \ \ \ \  for } j,k = L, R. \nonumber
\end{eqnarray}
In this coordinate system, the output wave-function in the outgoing far-field is expressed as
\begin{eqnarray}
&& \Psi_{{\rm out}}^{(jk)}(x_{j1},x_{k2}) \nonumber \\
&& {} \simeq \int^{\infty}_{-\infty} dx^{'}_{L1}dx^{'}_{L2} \nonumber \\
&& {} \times {{\bf u}}^{(jk)}_{{\rm 2ph}}(x_{j1},x_{k2};x^{'}_{L1},x^{'}_{L2}) \cdot \Psi_{{\rm in}}^{(LL)}(x^{'}_{L1},x^{'}_{L2}) \label{eq:simpletwophotonoutput} \\
&& {} \mbox{\ \ \ \ \ \ \ \ \ \ \ \ \ \ \ \ \ \ \ \ \ \ \ \ \ \ \ \ \ \ \ \ \ \ \ \ \ \ \ \ \ \ \ \ \ \ \ \ \ \ \ \  for } j,k = L, R. \nonumber
\end{eqnarray}
where \({\bf u}^{(jk)}_{{\rm 2ph}}(x_{j1},x_{k2};x^{'}_{L1},x^{'}_{L2})\) is given by
\begin{eqnarray}
{\bf u}^{(jk)}_{{\rm 2ph}}(x_{j1},x_{k2};x^{'}_{L1},x^{'}_{L2}) &=& {\bf u}^{{\rm lin} (jk)}_{{\rm 2ph}}(x_{j1},x_{k2};x^{'}_{L1},x^{'}_{L2}) \nonumber \\
&+& {\bf u}^{{\rm Nonlin} (jk)}_{{\rm 2ph}}(x_{j1},x_{k2};x^{'}_{L1},x^{'}_{L2}) \nonumber
\end{eqnarray}
\begin{eqnarray}
&& {} \mbox{ where } {\bf u}^{{\rm lin} (jk)}_{{\rm 2ph}}(x_{j1},x_{k2};x^{'}_{L1},x^{'}_{L2}) \nonumber \\
&& {} \ \ \ \ \ \ \ \ \ \ \ \ \ \ \ \ \ \ \ \ \ \ \ \ \ \ \ = {\bf u}_{{\rm 1ph}}^{(j)}(x_{j1};x^{'}_{L1}) \cdot {\bf u}_{{\rm 1ph}}^{(k)}(x_{k2};x^{'}_{L2}) \nonumber \\
&& \mbox{ with } \nonumber \\
&& {} \ \ {\bf u}_{\rm 1ph}^{(L)}(x_{L},x^{'}_{L}) = {\bf u}_{\rm abs}^{(L)}(x_{L},x^{'}_{L}) \nonumber \\
&& {} \ \ {\bf u}_{\rm 1ph}^{(R)}(x_{R},x^{'}_{L}) = {\bf u}_{\rm trans}^{(R)}(x_{R},x^{'}_{L}) + {\bf u}_{\rm abs}^{(R)}(x_{R},x^{'}_{L}) \nonumber
\end{eqnarray}
\begin{eqnarray}
&& {\bf u}_{\rm trans}^{(j)}(x_{j};x_{L}^{'}) = \delta \left( x^{'}_{L}-x_{j} \right) \nonumber \\
&& {\bf u}_{\rm abs}^{(j)}(x_{j};x_{L}^{'}) = -\frac{\Gamma}{2c} e^{-\frac{\Gamma}{2c} \left(x^{'}_{L}-x_{j}\right)} \mbox{\ for } x_{j} < x^{'}_{L}, \nonumber
\end{eqnarray}
\begin{eqnarray}
&& {} \mbox{ and } {\bf u}^{{\rm Nonlin} (jk)}_{{\rm 2ph}}(x_{j1},x_{k2};x^{'}_{L1},x^{'}_{L2}) \nonumber \\
&& {} \ \ \ \ \ \ \ \ \ \ \ \ \ \ = -{\bf u}_{\rm abs}^{(j)}(x_{j1},x^{'}_{L1}) \cdot {\bf u}_{\rm abs}^{(k)}(x_{k2},x^{'}_{L2}) \nonumber \\
&& {} \ \ \ \ \ \ \ \ \ \ \ \ \ \ \ \ \ \ \mbox{ for } 0 < x_{j1}, x_{k2} < \mbox{{\rm Min}} \left[x^{'}_{L1}, x^{'}_{L2} \right]. \nonumber
\end{eqnarray}

\section{Performance of QND measurement}
The performance of the proposed QND (Fig.~1) is evaluated using eq.~(\ref{eq:simpletwophotonoutput}). The pulsed mode of the one- and two-photon input is assumed to be a Gaussian mode \(\Psi_{{\rm in}}(x_{L}) = \sqrt{\frac{2}{d \sqrt{\pi}}} \exp \left[-2x_{L}^{2}/d^{2}\right]\), where $d$ is the input pulse duration. The one- and two-photon pulsed state can thus be described by
\begin{eqnarray}
&& \ket{\Psi^{{\rm 1ph}}_{{\rm in}}} = \int \ dx_{L} \Psi_{{\rm in}}(x_{L}) \ket{x_{L}} \mbox{ and} \\
&& \ket{\Psi^{{\rm 2ph}}_{{\rm in}}} = \ket{\Psi_{{\rm in 1}}} \otimes \ket{\Psi_{{\rm in 2}}} \nonumber \\
&& {} = \int \ dx_{L1}dx_{L2} \Psi_{{\rm in}}(x_{L1}) \cdot \Psi_{{\rm in}}(x_{L2}) \ket{x_{L1};x_{L2}}. \nonumber \\
&& 
\end{eqnarray}
The corresponding output states can be formulated using eq.~(\ref{eq:simpletwophotonoutput}) as
\begin{eqnarray}
&& \ket{\Psi^{{\rm 1ph}}_{{\rm out}}} = \sum_{j=L,R} \int \ dx_{L} \Psi^{(j)}_{{\rm out}}(x_{j}) \ket{x_{j}}\\
&& \ket{\Psi^{{\rm 2ph}}_{{\rm out}}} = \sum_{j,k=L,R} \int \ dx_{j1}dx_{k2} \Psi^{(jk)}_{{\rm out}}(x_{j1},x_{k2}) \ket{x_{j1};x_{k2}}, \nonumber \\
&&
\end{eqnarray}
where
\begin{eqnarray}
&& \Psi^{(j)}_{{\rm out}}(x_{j}) = \int \ dx{'}_{j} {\bf u}^{(j)}_{{\rm 1ph}}(x_{j};x^{'}_{L}) \cdot \Psi_{{\rm in}}(x^{'}_{L}) \nonumber \\
&& \Psi_{{\rm out}}^{(jk)}(x_{j1},x_{k2}) = \int \ dx^{'}_{L1} dx^{'}_{L2} \nonumber \\
&& \times {\bf u}^{(jk)}_{{\rm 2ph}}(x_{j1},x_{k2};x^{'}_{L1},x^{'}_{L2}) \cdot \Psi_{{\rm in}}(x^{'}_{L1}) \cdot \Psi_{{\rm in}}(x^{'}_{L2}) \nonumber
\end{eqnarray}
The transmittance and reflectance of the atom-cavity system for one- and two- photon input can be characterized by the detection probabilities on the left and right sides of the cavity, as given by
\begin{eqnarray}
&& {\rm P}_{{\rm 1ph}}(j;d) = \int dx_{j}\ \left| \Psi^{(j)}_{{\rm out}}(x_{j}) \right|^{2} \label{eq:proboneph} \\
&& {\rm P}_{{\rm 2ph}}(j1,k2;d) = \int dx_{j1}dx_{k2}\ \left| \Psi_{{\rm out}}^{(jk)}(x_{j1},x_{k2}) \right|^{2} \label{eq:probtwoph} \\
&& \ \ \ \ \ \ \ \ \ \ \ \ \ \ \ \ \ \ \ \ \ \ \ \ \ \ \ \ \ \ \ \ \mbox{for } j,k = L, R. \nonumber
\end{eqnarray}
As mentioned in the introduction, the performance of the QND can be characterized in terms of efficiency and success probability. The efficiency is defined as the probability that the signal photon appears at the output when the detector {\rm D1} or {\rm D2} (Fig.~\ref{fig:qnd}(b)) detects an ancillary photon. Note that the probability associated with detection of the arrival of the signal photon is different from the probability related to detection of the ancillary photon: the former is given by \({\rm P}_{{\rm 2ph}}(R1,L2;d) + {\rm P}_{{\rm 2ph}}(R1,R2;d)\), while the latter is given by adding \({\rm P}_{{\rm 1ph}}(R;d)\) to the former. The efficiency is thus given by the former divided by the latter, i.e.,
\begin{eqnarray}
&& {\rm EQND}(d) \nonumber \\
&& = \frac{{\rm P}_{{\rm 2ph}} (R1,R2;d) + {\rm P}_{{\rm 2ph}} (R1,L2;d)}{{\rm P}_{{\rm 1ph}}(R;d) +  {\rm P}_{{\rm 2ph}} (R1,R2;d) + {\rm P}_{{\rm 2ph}} (R1,L2;d)} \nonumber \\
&& {}
\end{eqnarray}
This equation is obtained using eqs.~(\ref{eq:proboneph}) and (\ref{eq:probtwoph}). Likewise, the success probability is defined as the probability that the detector {\rm D2} or {\rm D3} detects an ancillary photon when the signal photon appears at output 1 or 2 (Fig.~\ref{fig:qnd}(c)), as given by 
\begin{eqnarray}
&& {\rm P}_{suc}(d) = {\rm P}_{{\rm 2ph}} (R1,R2;d) + {\rm P}_{{\rm 2ph}} (R1,L2;d),
\end{eqnarray}
This equation holds when the state at the ancillary input port \({\rm A}_{in}\) on each path (Fig.~\ref{fig:qnd}(b)) is a single photon state. This condition gives the upper-limit of the success probability.

\begin{figure}[htbp]
\begin{picture}(0,0)
\put(-125,-10){(a)}
\put(-125, -220){(b)}
\end{picture}
\begin{center} 
\includegraphics[width=8.7cm]{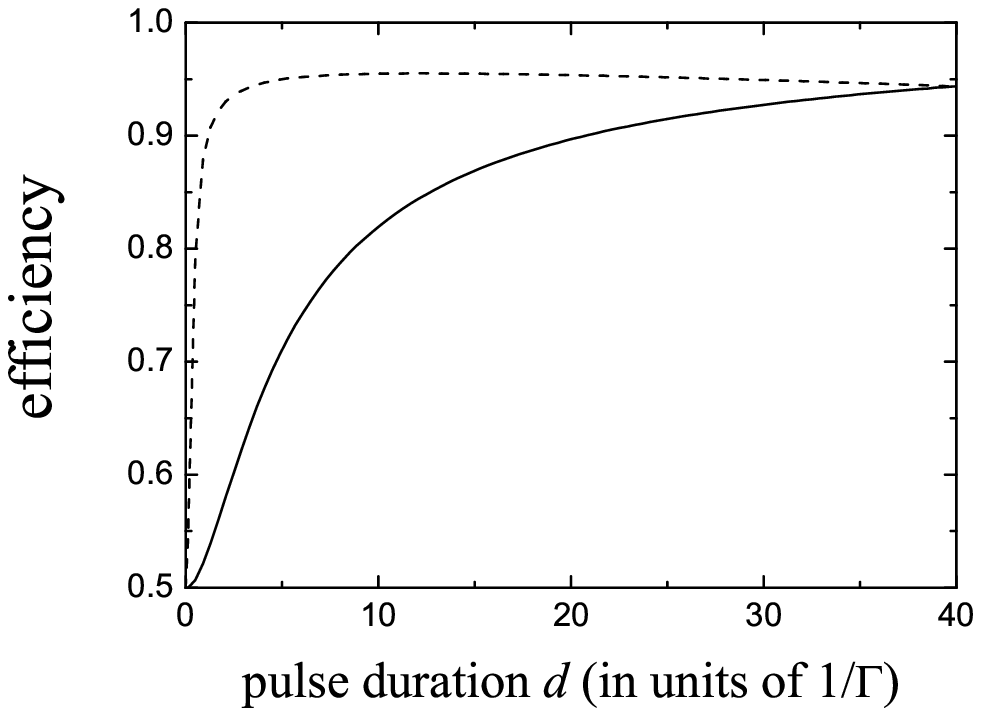}
\end{center}
\begin{center}
\includegraphics[width=8.7cm]{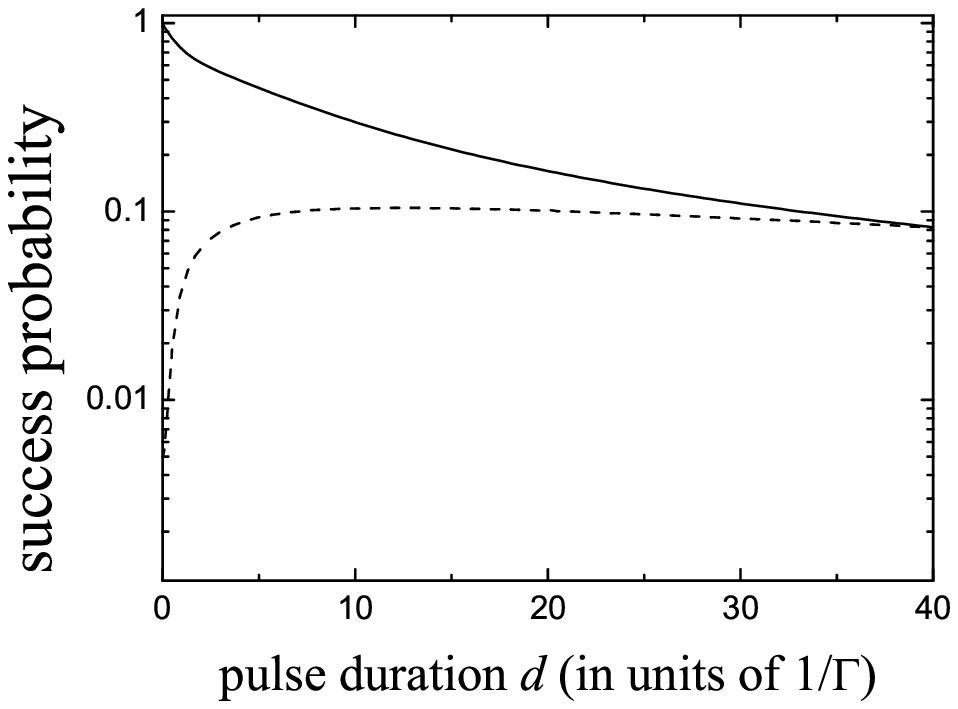}
\vspace{0.1cm}
\caption{\label{fig:effqnd} \scriptsize (a) Efficiency and (b) success probability of QND measurement. Solid and broken lines denote symmetric and asymmetric cases of pulse duration. The rate \(\Gamma\) is the dipole relaxation rate described by \(g^{2}/\kappa\), where \(g\) and \(\kappa\) represent the coupling constant of the atom-cavity system and the cavity decay rate, respectively}
\end{center} 
\end{figure}

Figure~\ref{fig:effqnd} shows the efficiency ${\rm EQND}(d)$ and success probability ${\rm P}_{suc}(d)$ for the proposed scheme. The symmetric case is that in which the pulse durations of the ancillary and signal photons are identical, while the asymmetric case is that in which the pulse duration of the ancillary photon is fixed at $40/\Gamma$, corresponding to an ancillary photon transmittance of $0.49 \%$. The efficiency can be increased to $100 \%$ by increasing the pulse duration of the ancillary photon. However, the success probability is simultaneously decreased to 0 \%. Considering this trade-off relationship, the efficiency at a success probability of $10\%$ is $94.3 \%$. A success probability of $8 \%$ is obtained for pulse durations of $40/\Gamma$ in the symmetric case. In the asymmetric case, efficiency of $95.5 \%$ was obtained with a success probability of $10 \%$ for pulse durations of $12.5/\Gamma$ (signal) and $40/\Gamma$ (ancillary). The asymmetric case thus relaxes the requirements for the signal pulse in the QND. For example, efficiency of greater than $90 \%$ is achievable over a wide range of signal photon pulse durations while maintaining a success probability of around $10 \%$. Furthermore, the jitter of the signal photon is larger than the pulse duration of the signal photon yet smaller than the pulse duration of the ancillary photon. The asymmetric case therefore allows the arrival time and pulse shape of the signal pulse to be determined by time-resolved photo-detection of the ancillary photon in conjunction with the output function eq.~(\ref{eq:simpletwophotonoutput}). Information on the pulse-shape of the output signal photon will become important when the signal photon is processed with another photon.

When the pulse durations of the signal and ancillary photons are much longer than the radiative relaxation time \(1/\Gamma\), the output wavefunctions associated with the detection of the ancillary photons, \(\Psi_{{\rm out}}^{(RL)}(x_{R1},x_{L2})\) and \(\Psi_{{\rm out}}^{(RR)}(x_{R1},x_{R2})\), can be approximated by the nonlinear component \(\int^{\infty}_{-\infty} dx^{'}_{1}dx^{'}_{2} {\bf u}^{{\rm Nonlin}}_{{\rm 2ph}}(x_{1},x_{2};x^{'}_{1},x^{'}_{2})\Psi_{{\rm in}}(x^{'}_{1}) \cdot \Psi_{{\rm in}}(x^{'}_{2})\). For simplicity, if the pulse shapes of the signal and ancillary photons are assumed to be rectangular and the pulse duration of the signal photon \(d_{2}\) is set much shorter than that of the ancillary photon \(d_{1}\), the nonlinear component can be expressed as \(-(1/d_{2})e^{-(\Gamma/2c)\left|x_{1}-x_{2}\right|}\). This function indicates that the pulse shape of the signal photon is not dependent on the detection timing of the ancillary photon. For a Gaussian pulse shape, the nonlinear component at \(x_{1}=x_{2}\) is described by a Gaussian function. However, the conclusion remains the same as in the rectangular case.

\section{Discussion}

One of the promising candidates for experimental realization of the two-sided atom-cavity is a single quantum dot exciton system coupled with the cavity mode of a photonic crystal \cite{englund}. Such a system provides design capabilities for temporal stability and reproductivity of the dipole coupling with the cavity mode. For a transition frequency and oscillator strength of the exciton of $\nu_{0}=2.35 \times 10^{5}$~GHz ($0.97$~eV) and $f=100$, the mode volume of a two-dimensional photonic crystal is $V_{m}=0.02~\mu$m$^{3}$, with a coupling constant $g$ of $0 \sim 132$~GHz. As the bad-cavity regime is typically $\kappa \simeq 4g$, the resultant radiative relaxation rate $\Gamma$ can take values of $0 \sim 33$~GHz. If the decoherence time of the exciton by phonons is ca. $1$~ns \cite{borri}, the pulse duration of the ancillary photon should be less than $500$~ps in order to avoid decoherence by phonons. Under this condition, the maximal efficiency for the QND is $86 \%$ and the success probability is $20 \%$ in the symmetric case. As there is little difference between the symmetric and asymmetric cases in terms of efficiency and success probability, the corresponding values for the asymmetric case should be similar. Note that the efficiency of $86 \%$ is a maximum, since the temporal evolution of a single exciton dipole under interaction with phonons, driven by weak coherent light with pulse duration of $500$~ps or more is unknown. This temporal evolution should therefore be investigated experimentally as part of future research. It will also be necessary to conduct detailed theoretical analyses of the decoherence time by phonons beyond the independent boson model \cite{badcavity, mahan} in order to discuss efficiencies of greater than $90 \%$.

\section{Conclusion}
A QND measurement scheme involving a two-sided atom-cavity system for the detection of photon arrival in entanglement sharing was proposed. The efficiency and success probability of the scheme were estimated by analyzing the responses of the two-sided atom-cavity system for one- and two-photon input over a range of input pulse duration. The conditions for improved QND performance were also examined. Efficiency of up to $100 \%$ was found to be possible by increasing the pulse duration of the ancillary photon, although the success probability is simultaneously reduced to $0 \%$ in a trade-off relationship. For a success probability of $10\%$, with relaxation of the requirements for the signal pulse in the QND, efficiency of $95.5 \%$ was obtained. In the case of signal photons with pulse duration of $12.5/\Gamma$ and ancilliary photons with pulse duration of $40/\Gamma$, the obtained success probability was $10 \%$. The success probability can be increased to $100\%$ by returning the signal photon to the input when no ancillary photon is detected on the right side of the cavity. Decoherence on the signal wave-packet after detection of the ancillary photon is suppressed by choosing signal and ancillary photon pulse durations much larger than the radiative relaxation time of the atom-cavity, in which case the pulse shape of the signal photon is approximated by the wave function \(\psi(x)=\sqrt{\aleph} \cdot e^{-(\Gamma/2c)\left|x\right|}\), where \(\aleph\) is a normalization factor.

The proposed QND scheme functions correctly even if the ancillary photon is replaced with weak light described by the superposition of vacuum and one-photon states. Therefore, the proposal is applicable not only for entanglement sharing but also for general purification of a single photon state. Realization of the proposal scheme is therefore expected to drive substantial progress in photon manipulation technology.


\appendix
\section{Derivation of eq.~(\ref{eq:intac})}
In the temporal evolution under the total Hamiltonian given by eq.~(\ref{eq:exthamiltonian}), the initial number of energy quanta is always preserved. For example, the number of energy quanta for a two-photon input is two, and this number is always preserved even upon interaction with the atom-cavity system. The possible states in the interaction of the V-type three-level system with two distinguishable photons {\rm Photon1} and {\rm Photon2} are thus
\begin{eqnarray}
&& {} \ket{{\rm g}} \otimes \ket{{\rm F}^{(l)}_{1}} \otimes \ket{{\rm F}^{(l^{'})}_{2}} \nonumber \\
&& {} \ket{\xi_{1}} \otimes \ket{{\rm F}^{(0)}_{1}} \otimes \ket{{\rm F}^{(l^{'})}_{2}} \nonumber \\
&& {} \ket{\xi_{2}} \otimes \ket{{\rm F}^{(l)}_{1}} \otimes \ket{{\rm F}^{(0)}_{2}} \ \ \ \mbox{for } l,l^{'} = 1, 2, 3 \nonumber \\
&& {} \mbox{, where } \ket{{\rm F}^{(0)}_{m}} \equiv \ket{0_{a_{m}}, 0_{k_{Lm}}, 0_{k_{Rm}}}, \nonumber \\
&& {} \ \ \ \ \ \ \ \ \ \ \ \ket{{\rm F}^{(1)}_{m}} \equiv \ket{1_{a_{m}}, 0_{k_{Lm}}, 0_{k_{Rm}}}, \nonumber \\
&& {} \ \ \ \ \ \ \ \ \ \ \  \ket{{\rm F}^{(2)}_{m}} \equiv \ket{0_{a_{m}}, 1_{k_{Lm}}, 0_{k_{Rm}}} \nonumber \\
&& {} \ \ \ \mbox{, and } \ket{{\rm F}^{(3)}_{m}} \equiv \ket{0_{a_{m}}, 0_{k_{Lm}}, 1_{k_{Rm}}}. \nonumber
\end{eqnarray}
The state \(\ket{0_{a_{m}}, 1_{k_{Lm}}, 0_{k_{Rm}}}\) denotes a state in which the cavity mode \(a_{m}\) and all modes of the "Rm field" \(k_{Rm}\) are in the vacuum state, and one mode of the "Lm field" \(k_{Lm}\) is in the first excited state, with the remaining states being the vacuum state. The state \(\ket{0_{a_{m}}, 0_{k_{Lm}}, 0_{k_{Rm}}}\) denotes a state in which the cavity mode \(a_{m}\) and all modes of the "Rm field" \(k_{Rm}\) and the "Lm field" \(k_{Lm}\) are in the vacuum state. The same holds for \(\ket{1_{a_{m}}, 0_{k_{Lm}}, 0_{k_{Rm}}}\) and \(\ket{0_{a_{m}}, 0_{k_{Lm}}, 1_{k_{Rm}}}\).

On the truncated Hilbert space composed of these states, the matrix representation of the operators \(\hat{a}_{1}^{\dagger} \hat{\sigma}^{(1)}_{-}\) and \(\hat{a}_{2}^{\dagger} \hat{\sigma}^{(2)}_{-}\) are given by
\begin{eqnarray}
\hat{a}_{1}^{\dagger} \hat{\sigma}^{(1)}_{-} = && {} \ketbra{{\rm g}}{\xi_{1}} \otimes  \ketbra{{\rm F}^{(1)}_{1}}{{\rm F}^{(0)}_{1}} \nonumber \\
&& {} \otimes \left( \ketbra{{\rm F}^{(1)}_{2}}{{\rm F}^{(1)}_{2}} + \int^{\infty}_{-\infty} dk_{L2} \ \ketbra{{\rm F}^{(2)}_{2}}{{\rm F}^{(2)}_{2}} \right. \nonumber \\
&& {} \ \ \ \ \ \ \ \ \ \ \ \ \ \ \ \ \ \ \ \ \left. + \int^{\infty}_{-\infty} dk_{R2} \ \ketbra{{\rm F}^{(3)}_{2}}{{\rm F}^{(3)}_{2}} \right) \nonumber \\
&& {} \label{eq:truncatedone}\\
\hat{a}_{2}^{\dagger} \hat{\sigma}^{(2)}_{-} = && {} \ketbra{{\rm g}}{\xi_{2}} \nonumber \\
&& {} \otimes \left( \ketbra{{\rm F}^{(1)}_{1}}{{\rm F}^{(1)}_{1}} + \int^{\infty}_{-\infty} dk_{L1} \ \ketbra{{\rm F}^{(2)}_{1}}{{\rm F}^{(2)}_{1}} \right. \nonumber \\
&& {} \ \ \ \ \ \ \ \ \ \ \ \ \ \ \ \ \ \ \ \ \left. + \int^{\infty}_{-\infty} dk_{R1} \ \ketbra{{\rm F}^{(3)}_{1}}{{\rm F}^{(3)}_{1}} \right) \nonumber \\
&& {} \otimes  \ketbra{{\rm F}^{(1)}_{2}}{{\rm F}^{(0)}_{2}}. \nonumber \\
&& {} \label{eq:truncatedtwo}
\end{eqnarray}
To express the above operators as those acting on the Hilbert space spanned by the state descriptions given by eq.~(\ref{eq:twophotondescription}), we start from expressing the quantum state of the V-type atomic system as the quantum state of the two two-level atomic systems where the Hilbert space is spanned by the basis \{\(\ket{{\rm g}_{1},{\rm g}_{2}}, \ket{{\rm e}_{1},{\rm g}_{2}}, \ket{{\rm g}_{1},{\rm e}_{2}}, \ket{{\rm e}_{1},{\rm e}_{2}}\)\}. The V-type atomic system, where the excited state is \(\ket{\xi_{1}}\) or \(\ket{\xi_{2}}\) or the superposition of these states, doesn't emit two photons simultaneously. On the other hand, the two two-level atomic systems in the double excited state (\(\ket{{\rm e}_{1},{\rm e}_{2}}\)) emit two photons simultaneously. This difference implies that the quantum state of the V-type system should be expressed as a quantum state in the subspace spanned by the basis except for the double excited state \(\ket{{\rm e}_{1},{\rm e}_{2}}\) of the two two-level atomic systems. The ground state of the V-type atomic system \(\ket{{\rm g}}\) corresponds to the ground state of the two two-level atomic systems \(\ket{{\rm g}_{1},{\rm g}_{2}}\). From the view-point of single photon resonant transition processes, the allowed transition to and from either of the two excited states \(\ket{\xi_{1}}\) and \(\ket{\xi_{2}}\) correspond to the transition to and from either of the two excited states \(\ket{{\rm e}_{1},{\rm g}_{2}}\) and \(\ket{{\rm g}_{1},{\rm e}_{2}}\). Using these correspondences, the operator \(\ketbra{{\rm g}}{\xi_{1}}\) in eq.~(\ref{eq:truncatedone}) is expressed as the operator \(\ketbra{{\rm g}_{1},{\rm g}_{2}}{{\rm e}_{1},{\rm g}_{2}}=\ketbra{{\rm g}_{1}}{{\rm e}_{1}} \otimes \ketbra{{\rm g}_{2}}{{\rm g}_{2}}\) on the Hilbert space of the two two-level atomic systems. Likewise, The operator \(\ketbra{{\rm g}}{\xi_{2}}\) in eq.~(\ref{eq:truncatedtwo}) is expressed as the operator \(\ketbra{{\rm g}_{1}}{{\rm g}_{1}} \otimes \ketbra{{\rm g}_{2}}{{\rm e}_{2}}\). The operators given by eqs.~(\ref{eq:truncatedone}) and (\ref{eq:truncatedtwo}) are then expressed by substituting these expressions as
\begin{eqnarray}
\hat{a}_{1}^{\dagger} \hat{\sigma}^{(1)}_{-} = && {} \ketbra{{\rm g}_{1}}{{e}_{1}} \otimes  \ketbra{{\rm F}^{(1)}_{1}}{{\rm F}^{(0)}_{1}} \otimes \ketbra{{\rm g}_{2}}{{\rm g}_{2}} \nonumber \\
&& {} \otimes \left( \ketbra{{\rm F}^{(1)}_{2}}{{\rm F}^{(1)}_{2}} + \int^{\infty}_{-\infty} dk_{L2} \ \ketbra{{\rm F}^{(2)}_{2}}{{\rm F}^{(2)}_{2}} \right. \nonumber \\
&& {} \ \ \ \ \ \ \ \ \ \ \ \ \ \ \ \ \ \ \ \ \left. + \int^{\infty}_{-\infty} dk_{R2} \ \ketbra{{\rm F}^{(3)}_{2}}{{\rm F}^{(3)}_{2}} \right) \nonumber \\
&& {} \\
\hat{a}_{2}^{\dagger} \hat{\sigma}^{(2)}_{-} = && {} \ketbra{{\rm g}_{1}}{{\rm g}_{1}} \nonumber \\
&& {} \otimes \left( \ketbra{{\rm F}^{(1)}_{1}}{{\rm F}^{(1)}_{1}} + \int^{\infty}_{-\infty} dk_{L1} \ \ketbra{{\rm F}^{(2)}_{1}}{{\rm F}^{(2)}_{1}} \right. \nonumber \\
&& {} \ \ \ \ \ \ \ \ \ \ \ \ \ \ \ \ \ \ \ \ \left. + \int^{\infty}_{-\infty} dk_{R1} \ \ketbra{{\rm F}^{(3)}_{1}}{{\rm F}^{(3)}_{1}} \right) \nonumber \\
&& {} \otimes  \ketbra{{\rm g}_{2}}{{\rm e}_{2}} \otimes \ketbra{{\rm F}^{(1)}_{2}}{{\rm F}^{(0)}_{2}}
\end{eqnarray}
These operators can be expressed on the state descriptions \(\ket{k_{Lj}}\) \(\ket{k_{Rj}}\), \(\ket{{\rm C}_{j}}\), and \(\ket{{\rm E}_{j}}\) for \(j=1,2\) as
\begin{eqnarray}
&& {} \hat{a}_{1}^{\dagger} \hat{\sigma}^{(1)}_{-} = \ketbra{{\rm C}_{1}}{{\rm E}_{1}} \nonumber \\
&& {} \ \ \ \ \ \ \ \ \ \otimes \left(\ketbra{{\rm C}_{2}}{{\rm C}_{2}} + \sum_{i=L,R} \ \int^{\infty}_{-\infty} dk_{i2} \ \ketbra{k_{i2}}{k_{i2}}\right) \nonumber \\
&& {} \\
&& {} \hat{a}_{2}^{\dagger} \hat{\sigma}^{(2)}_{-} = \left(\ketbra{{\rm C}_{1}}{{\rm C}_{1}} + \sum_{i=L,R} \ \int^{\infty}_{-\infty} dk_{i1} \ \ketbra{k_{i1}}{k_{i1}}\right) \nonumber \\
&& {} \ \ \ \ \ \ \ \ \ \otimes \ketbra{{\rm C}_{2}}{{\rm E}_{2}}. \nonumber \\
&& {} 
\end{eqnarray}
The matrix representation of the interaction Hamiltonian \(\sum_{j=1,2} \ \hat{H}^{(j)}_{intac}\) given by eq.~(\ref{eq:intac}) is thus obtained by the above equations.

\section*{Acknowledgement}
K.K thanks NEC researchers Mr.~Sirane and Mr.~Kirihara for giving valuable information about the current state on the study of photonic crystal cavity and sharing entangled atomic systems.

\end{document}